\documentclass[twocolumn,prd,unsortedaddress,superscriptaddress,showpacs,a4paper,nofootinbib]{revtex4-1}
\usepackage[T1]{fontenc}
\usepackage{graphicx}
\usepackage{indentfirst}
\usepackage{pdfpages}
\usepackage{multirow}
\usepackage{amssymb}
\usepackage{amsfonts}
\usepackage{amsmath}
\usepackage{epsfig}

\def\beq{\begin{equation}}
\def\enq{\end{equation}}
\def\beqa{\begin{eqnarray}}
\def\enqa{\end{eqnarray}}

\def\GeV{\nobreak\,\mbox{GeV}}

\def\qq{\lag\bar{q}q\rag}

\def\mix{\lag\bar{q}g\si.Gq\rag}

\def\qGq{\lag\bar{q}Gq\rag}
\def\GG{\lag G^2 \rag}
\def\gG{\lag g_s^2 G^2 \rag}
\def\G3{\lag g^3G^3\rag}
\def\pli{p^\prime}

\def\la{\lambda}

\def\ga{\gamma}
\def\Ga{\Gamma}

\def\si{\sigma}

\def\al{\alpha}
\def\be{\beta}
\def\almax{\alpha_{max}}
\def\almin{\alpha_{min}}
\def\bemax{1-\al}
\def\bemin{\beta_{min}}
\def\lb{\label}
\def\nn{\nonumber}

\newcommand{\rag}{\rangle}
\newcommand{\lag}{\langle}

\begin{document}
\title{$Y(4260)$ as a mixed charmonium-tetraquark state}
\author{J.M. Dias}
\email{jdias@if.usp.br}
\author{R.M. Albuquerque}
\email{rma@if.usp.br}
\author{M. Nielsen}
\email{mnielsen@if.usp.br}
\affiliation{Instituto de F\'{\i}sica, Universidade de S\~{a}o Paulo, 
C.P. 66318, 05389-970 S\~{a}o Paulo, SP, Brasil}
\author{C.M. Zanetti}
\email{carina.zanetti@gmail.com}
\affiliation{ Faculdade de Tecnologia, Universidade do Estado do Rio de 
Janeiro, Rod. Presidente Dutra Km 298, P\'olo Industrial, 27537-000 , 
Resende, RJ, Brasil}

\begin{abstract}
Using the QCD sum rule approach we study the $Y(4260)$ state assuming that it 
can be described by a mixed charmonium-tetraquark current with 
$J^{PC}=1^{--}$ quantum numbers. For the mixing angle around $\theta \approx 
(53.0\pm 0.5)^{0}$, we obtain a value for the mass which is
in good agreement with the experimental mass  of the $Y(4260)$. 
However, for the decay width we find the value $\Ga_Y \approx
(1.0\pm 0.2)$ MeV which is not compatible with the experimental 
value $\Ga \approx (88\pm 23)$ MeV. Therefore, we conclude
that, although we can explain the mass of the $Y(4260)$, this state 
cannot be described as a mixed charmonium-tetraquark state since, with
this assumption, we can not explain its decay width.
\end{abstract}

\maketitle
\section{Introduction}

Many of the charmonium-like states recently observed in  $e^+e^-$ collisions 
by BaBar and Belle collaborations do not fit the quarkonia interpretation,
 and have stimulated an extensive discussion about exotic hadron
configurations. The production mechanism, masses, decay widths, spin-parity 
assignments and decay modes of these states, called  $X,~Y$ and $Z$ states,
have been discussed in some reviews  
\cite{Swanson:2006st,Zhu:2007wz,Nielsen:2009uh,Olsen:2009gi,Brambilla:2010cs}).
Among these states, the $Y(4260)$ is particularly interesting. It was first
observed by BaBar collaboration in the $e^+e^-$ annihilation through 
initial state radiation \cite{babar1}, and it was confirmed by CLEO and Belle 
collaborations \cite{yexp}. The $Y(4260)$ was also observed in the 
$B^-\to Y(4260)K^-\to J/\Psi\pi^+\pi^-K^-$ decay \cite{babary2}, and CLEO
reported two additional decay channels: $J/\Psi\pi^0\pi^0$ and
$J/\Psi K^+K^-$ \cite{yexp}. 

Since the mass of the $Y(4260)$ is higher than the $D^{(*)}\bar{D}^{(*)}$
threshold, if it was a normal $c\bar{c}$ charmonium state, it should decay 
mainly to $D^{(*)}\bar{D}^{(*)}$. However, the
observed $Y$ state do not match the peaks in $e^+e^-\to D^{(*)\pm}D^{(*)
\mp}$ cross sections measured  by Belle \cite{belle5} and BaBar 
\cite{babar5,babar6}.
Besides, the $\Psi(3S),~\Psi(2D)$ and $\Psi(4S)$ $c\bar{c}$ states have 
been assigned to the well established $\Psi(4040),~\Psi(4160),~$ and 
$\Psi(4415)$ mesons respectively, and the prediction from quark models 
for the $\Psi(3D)$  state is 4.52 GeV. Therefore, the mass of the $Y(4260)$
is not consistent with any of the $1^{--}$ $c\bar{c}$ states 
\cite{Zhu:2007wz,Nielsen:2009uh,kz}. 

There are many theoretical interpretations for 
the $Y(4260)$: tetraquark state \cite{tetraquark}, hadronic molecule of 
$D_{1} D$, $D_{0} D^*$ \cite{Ding}, $\chi_{c1} \omega$ 
\cite{Yuan}, $\chi_{c1} \rho$ \cite{liu}, $J/\psi f_0(980)$ \cite{oset}, 
a hybrid charmonium \cite{zhu}, a charm baryonium \cite{Qiao}, a cusp 
\cite{eef1,eef2,eef3}, etc. Within the available experimental information, 
none of these suggestions can be completely ruled out. However, there are some
calculations, within the QCD sum rules (QCDSR) approach
\cite{Nielsen:2009uh,svz,rry,SNB}, that can not explain 
the mass of the $Y(4260)$ supposing it to be a tetraquark state \cite{rapha}, 
or  a $D_{1} D$, $D_{0} D^*$ hadronic molecule \cite{rapha}, or  a  
$J/\psi f_0(980)$ molecular state \cite{Albuquerque:2011ix}. 

In this work we 
use again the QCDSR approach to the $Y(4260)$ state including a new 
possibility: the mixing between two and four-quark states. This will be 
implemented  following the prescription suggested in \cite{oka24} for the light 
scalar mesons. The mixing is done at the level of the currents and was 
extended to the charm sector in Ref. \cite{matheus}, in order to study
the $X(3872)$ as a mixed charmonium-molecular state. In particular, in 
Ref. \cite{matheus}, the mass and the  decay width of the $X(3872)$, into 
$2\pi$ and $3\pi$, were evaluated  with good agreement with the experimental 
values. Agreement with the experimental results has been also obtained, 
applying this same approach, in the study of the $X(3872)$ radiative decay 
\cite{x3872rad}, and also in the $X(3872)$ production rate in $B$ decay
\cite{x3872prod}.

In the next sections we  consider a mixed charmonium-tetraquark current and
use the QCDSR method to study both, mass and decay width, of the $Y(4260)$.

%
\section{Constructing the two-quark and four-quark operator}
%

In order to define a mixed charmonium-tetraquark current 
we have to define the currents associated with
charmonium and four-quarks (tetraquark) states. For the
charmonium part we use the conventional vector current:
\beq
j_\mu^{'(2)}=\bar{c}_a(x)\ga_\mu c_a(x),
\label{jcc}
\enq
while the tetraquark part is interpolated by \cite{rapha}
\beqa
j_\mu^{(4)} &=& \frac{\epsilon_{abc} \epsilon_{dec}}{\sqrt{2}}
\Big[(q_a^T(x)C\ga_5 c_b(x))(\bar{q}_d(x)\ga_\mu\ga_5 C\bar{c}_e^T(x))+\nn\\
&&+(q_a^T(x)C\ga_5\ga_\mu c_b(x))(\bar{q}_d(x)\ga_5 C\bar{c}_e^T(x)) \Big].
\label{j4q}
\enqa
As in Refs. \cite{oka24,matheus}, we define the normalized two-quark
current as
\beq
j_\mu^{(2)}=\frac{1}{\sqrt{2}}\qq ~j_\mu^{'(2)},
\enq
and from these two currents we build the following mixed
charmonium-tetraquark $J^{PC}=1^{--}$ current for the $Y(4260)$ state:

\beq
j_\mu(x)=\sin(\theta) \:j_\mu^{(4)}(x)+\cos(\theta) \:j_\mu^{(2)}(x),
\label{jmix}
\enq

%
\section{The Two-Point Correlation Function}
%

To obtain the mass of a hadronic state using the QCDSR
approach, the starting point is the two-point correlation function
\begin{align}
\Pi_{\mu\nu}(q) &= i \int d^{4}x ~e^{i q\cdot x}
\langle 0| \:T[ j_{\mu}(x) j_{\nu}^{\dagger} (0) ]
 \:|0 \rangle \nonumber \\
 &= -\Pi_{1}(q^{2})\Big(g_{\mu \nu} - \frac{q_{\mu}q_{\nu}}{q^{2}}
 \Big) + \Pi_{0}(q^{2})
 \frac{q_{\mu}q_{\nu}}{q^{2}},
\label{2point}
\end{align}
where $j_{\mu}(x)$ is the mixed charmonium-tetraquark
interpolating current defined in Eq.~(\ref{jmix}). 
The functions $\Pi_{1}(q^{2})$ and $\Pi_{0}(q^{2})$ are 
two independent invariant functions associated with spin-1 
and spin-0 mesons, respectively.

According to the principle of duality, Eq. (\ref{2point})
can be evaluated in two ways: in the OPE side, we calculate 
the correlation function in terms of quarks and gluon fields 
using the Wilson's operator product expansion. 
The phenomenological side   is evaluated by inserting, in 
Eq. (\ref{2point}), a complete set of intermediate states with $1^{--}$ 
quantum numbers. In this side, we parametrize the 
coupling of the vector state $Y$ with the current defined 
in Eq. (\ref{jmix}) through the coupling 
parameter $\lambda_Y$
\begin{equation}
\langle 0| j_{\mu}(x)|Y\rangle = \lambda_Y \epsilon_{\mu}.
\label{coupling}
\end{equation}
where $\epsilon_\mu$ is the polarization vector. Using 
Eq. (\ref{coupling}), we can write the phenomenological
side of Eq. (\ref{2point}) as

\begin{equation}
\Pi^{fen}_{\mu \nu}(q) = \frac{\lambda_Y^{2}}{M_{Y}^{2} - q^{2}}
\Big(g_{\mu \nu} - \frac{q_{\mu}q_{\nu}}
{q^{2}} \Big) + \:.\:.\:.\:
\label{phenoside}
\end{equation}
where $m_Y$ is the mass of the $Y$ state and the dots, in the 
second term in the RHS of Eq. (\ref{phenoside}), denotes the 
higher resonance contributions which will be parametrized, as usual, 
through introduction of the continuum threshold parameter $s_{0}$ 
\cite{io1}.

The OPE side can be written in terms of a dispersion relation
\begin{equation}
\Pi^{OPE}(q^{2}) = \int\limits_{4m^{2}_c}^{\infty}
 ds\frac{\rho^{OPE}(s)}{s - q^{2}},
\end{equation}
where $\rho^{OPE}(s)$ is given by the imaginary part 
of the correlation function: $\pi \rho^{OPE}(s) = Im[\Pi^{OPE}(s)]$. 
In this side, we work at leading order in $\alpha_{s}$ in the 
operators and we consider the contributions from the
condensates up to dimension eight. Although we will consider only a 
part of the of the dimension 8 condensates (related to the quark 
condensate times the mixed condensate), in Ref.~\cite{finazzo} it was shown
that this is the most important dimension 8 condensate contribution.

Considering the current in Eq.~(\ref{jmix}),
Eq. (\ref{2point}) in the OPE side can be written as
\begin{align}
\Pi_{\mu \nu}(q) &= \frac{\langle \bar{q}q\rangle^{2}}{2} 
\:cos^{2}\:(\theta) ~\Pi^{22}_{\mu \nu}(q) + sin^{2}\:(\theta)
 ~\Pi^{44}_{\mu \nu}(q) \nonumber \\  
&+ \frac{\langle \bar{q}q\rangle}{\sqrt{2}} \:sin\:(\theta)
 \:cos\:(\theta) \bigg[ \Pi^{24}_{\mu \nu}(q)
+ \Pi^{42}_{\mu \nu}(q) \bigg] ,
\label{opeside}
\end{align}
with
\begin{equation}
\Pi_{\mu \nu}^{ij}(q) = i \int d^{4}x ~e^{i q\cdot x}
\langle 0|T [j_{\mu}^{i}(x)j_{\nu}^{j\dagger}(0)
]|0\rangle .
\end{equation}
Clearly $\Pi_{\mu \nu}^{22}(q)$ and $\Pi_{\mu \nu}^{44}(q)$
are, respectively, the correlation functions of the $J/\psi$ 
and $[cq][\bar{c}\bar{q}]$ tetraquark state.

After making a Borel transform in both sides, and 
transferring the continuum contributions to the 
OPE side, the sum rule in the $g_{\mu \nu}$ structure 
for the vector meson can be written as
\begin{align}
\lambda_Y^{2}e^{- m_{Y}^{2}/M_B^{2}} &= 
\frac{\langle \bar{q}q\rangle^{2}}{2}cos^{2}(\theta)
~\Pi^{22}_{1}(M_B^{2}) + sin^{2}(\theta) 
~\Pi^{44}_{1}(M_B^{2}) \nonumber \\  
&+ \frac{\langle \bar{q}q\rangle}{\sqrt{2}}
 sin(\theta) cos(\theta) \bigg[ \Pi^{24}_{1}(M_B^{2}) + 
\Pi^{42}_{1}(M_B^{2}) \bigg],
\label{sumrule}
\end{align}
where
\begin{align}
\Pi^{22}_{1}(M_B^{2}) &= \int\limits^{s_0}_{4m_{c}^{2}}
\: ds ~e^{-s/M_B^{2}} \rho^{22}_{pert}(s)
~+~ \Pi_{\GG}^{22}(M_B^{2}),  \nonumber\\
\label{pi22}
\end{align}
\begin{align}
\Pi^{44}_{1}(M_B^{2}) &= \int\limits^{s_0}_{4m_{c}^{2}}
\: ds ~e^{-s/M_B^{2}}\Bigg( \rho^{44}_{pert}(s) +
\rho^{44}_{\qq}(s) + \nn\\ 
&+\rho^{44}_{\GG}(s) + \rho^{44}_{\qGq}(s) + 
\rho^{44}_{\qq^2}(s) +\nn\\
&+\rho^{44}_{\lag 8 \rag} \Bigg) + 
\Pi_{\lag 8 \rag}^{44}(M_B^{2}),
\label{pi44}
\end{align}
\begin{align}
\Pi^{24}_{1}(M_B^{2}) &= \int\limits^{s_0}_{4m_c^2}
\: ds ~e^{-s/M_B^{2}} \rho_{\qq}^{24}(s)
+ \Pi_{\qGq}^{24}(M_B^{2}).
\label{pi24}
\end{align}

The expressions for the spectral density $\rho(s)$ 
appearing in Eqs. (\ref{pi22}) - (\ref{pi24}) for 
the charmonium and tetraquark states, as well as the mixed 
terms are listed in  Appendix A.

By taking the derivative of Eq. (\ref{sumrule}) with 
respect to $1/M_B^{2}$ and dividing the result by Eq. 
(\ref{sumrule}), we obtain
\begin{equation}
m_{Y}^{2} = -\frac{\frac{dK(M_B^{2},\theta)}
{d(1/M_B^{2})}}{K(M_B^{2},\theta)},
\label{mass}
\end{equation}
where 
\begin{eqnarray}
K(M_B^{2},\theta) &\equiv \frac{\langle \bar{q}q\rangle^{2}}
{2}cos^{2}(\theta)\Pi^{22}_{1}(M_B^{2}) + sin^{2}(\theta) 
\Pi^{44}_{1}(M_B^{2}) \nonumber \\  
&+ \frac{\langle \bar{q}q\rangle}{\sqrt{2}} sin(\theta)
 cos(\theta) \bigg[ \Pi^{24}_{1}(M_B^{2}) + 
\Pi^{42}_{1}(M_B^{2}) \bigg]\nn.
\end{eqnarray}
Eq. (\ref{mass}) will be used to extract 
the mass of the charmonium-tetraquark state.

%
\subsection{Numerical Analysis}
%

In Table \ref{Param} we list the values of the quark masses and condensates
that we have used in our numerical analysis. For a 
consistent comparison with results obtained for the 
others works using QCD sum rules, these parameters 
values used here are the same values used in 
Refs. \cite{x3872,SNB,Albuquerque:2011ix,narpdg}.

{\small
\begin{table}[h]
\setlength{\tabcolsep}{1.25pc}
\caption{Quark masses and condensates values.}
\begin{tabular}{ll}
&\\
\hline
Parameters&Values\\
\hline
%
$m_{c}(m_{c})$ & $(1.23 \pm 0.05) ~\mbox{GeV}$ \\
$\langle \bar{q}q \rangle$ & $-(0.23 \pm 0.03)^{3} ~\mbox{GeV}^{3}$\\
$\langle \bar{q}g\sigma.Gq \rangle$ & $m_{0}^{2}\langle \bar{q}q\rangle$\\
$m_{0}^{2} $ & $(0.8 \pm 0.1)~\mbox{GeV}^{2}$\\
$\langle g_{s}^{2} G^{2} \rangle$ & $(0.88\pm 0.25)~\mbox{GeV}^{4}$\\
\hline
\end{tabular}
\label{Param}
\end{table}}

The continuum threshold is a physical parameter that, in the QCDSR approach, 
should be related to the first excited state with the same quantum 
numbers. In some known cases, like 
the $\rho$ and $J/\psi$, the first excited state has
a mass approximately $0.5$ GeV above the ground state mass. 
Since in our study we do not know the experimental spectrum for the 
hadrons studied, we will fix the continuum threshold range starting with 
the smaller value which provides a valid Borel window, as explained below.
Using this criterion, we obtain $s_{0}$ in the
range $4.6 \leq \sqrt{s_{0}} \leq 4.8$ GeV.

Reliable results can be extracted from the sum rule if is possible 
to determine a valid Borel Window. Such Borel window is obtained by 
imposing a good  OPE convergence, the dominance of the pole contribution and 
a good Borel stability. 
To determine the minimum value of the Borel mass we adopt the 
criterion for which the contribution of the higher dimension 
condensate should be smaller than 15\% of the total contribution. 
Thus, $M_{Bmin}^{2}$ is such that
\begin{eqnarray}
\Bigg|\frac{\mbox{OPE summed up to dim n-1}
 (M_{Bmin}^{2})}{\mbox{total contribution} 
 (M_{Bmin}^{2})}\Bigg| = 0.85.\nn \\
\end{eqnarray}

\begin{figure}[h] 
\begin{center}
\centerline{\includegraphics[width=0.48\textwidth]{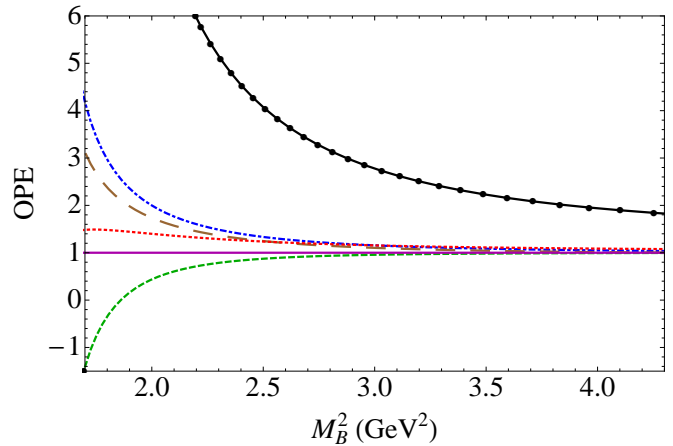}}
\caption{The OPE convergence in the region 
$2.0 \leq M^{2}_{B} \leq 6.0$ GeV$^{2}$ for
$\sqrt{s_{0}} =4.70$ GeV. We plot the relative contributions
start with perturbative contribution (line with circles), 
and each other lines represents the relative contribution after 
adding of one extra condensate in expansion: $+ \langle \bar{q}q\rangle$
(dot-dashed line), $+ \langle G^{2}\rangle$ (long-dashed line), 
$+ \langle \bar{q}g\sigma.Gq\rangle$ (dotted line), $+\qq^2$ 
(dashed line) and $\qq\mix$ (solid line).}
\label{fig1} 
\end{center}
\end{figure}

In Fig. \ref{fig1} we plot the relative contributions 
of all the terms in the OPE side. We have used 
$\sqrt{s_{0}} = 4.70$ GeV and $\theta = 53^{0}$. For 
others $\theta$ values outside the range $52.5^{0}\leq\theta\leq53.5^0$,
we do not have a good OPE convergence.
From this figure we see that the contribution 
of the dimension-8 condensates is smaller 
than 15\% of the total contribution for values of 
$M^{2}_{B} \geq 2.4$ GeV$^{2}$, indicating a good OPE 
convergence. Therefore, we fix the lower value of
$M^{2}_{B}$ in the sum rule window as: 
$M_{Bmin}^{2} = 2.4$ GeV$^{2}$.

To determine the maximum value of the Borel mass
($M_{Bmax}^{2}$) we must analyse the pole-continuum
contribution. Unlike the pole contribution, the continuum 
contribution increases with $M^{2}_{B}$ due to
the dominance of the perturbative contribution. Therefore, the maximum 
value of the Borel mass is determined in the point that the
pole contribution is equal to the continuum contribution.

\begin{figure}[h]
\begin{center}
\centerline{\includegraphics[width=0.48\textwidth]{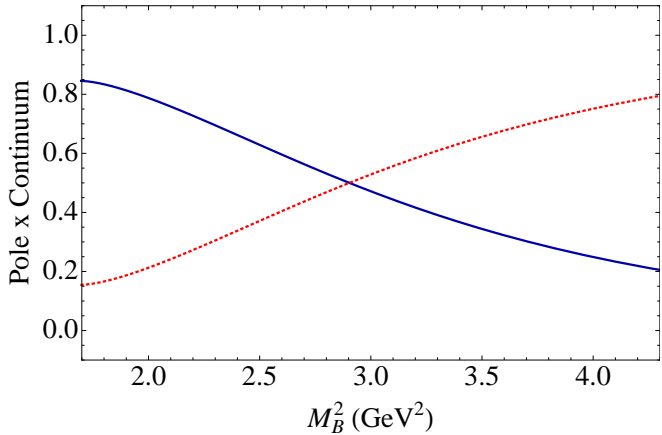}}
\caption{The pole contribution (divided by the total, pole 
plus continuum, contribution) represented by solid line
and the continuum contribution (dotted line) for the
$\sqrt{s_{0}} =4.70$ GeV.}
\label{fig2} 
\end{center}
\end{figure}  

In Fig. \ref{fig2} we see a comparison between 
the pole and continuum contributions. It is clear that 
the pole contribution is equal to the continuum 
contribution for $M^{2}_B = 2.90$ GeV$^{2}$. Therefore, for 
$\sqrt{s_{0}} = 4.70$ GeV$^{2}$ and $\theta = 53^{0}$ the 
Borel window is: $2.4 \leq M^{2}_{B} \leq 2.90$
GeV$^{2}$. 

\begin{figure}[h]
\begin{center}
\centerline{\includegraphics[width=0.48\textwidth]{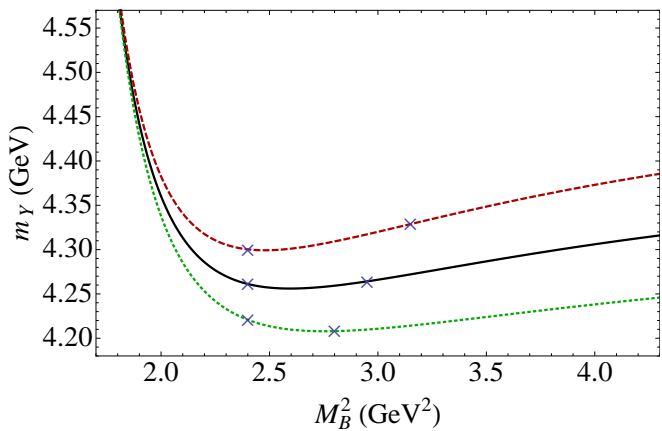}}
\caption{The mass as a function of the sum rule parameter 
$M^{2}_{B}$ for $\sqrt{s_{0}} = 4.60$ GeV (dotted line), 
$\sqrt{s_{0}} = 4.70$ GeV (solid line), $\sqrt{s_{0}} = 4.80$ GeV
(long-dashed line). The crosses indicate the valid Borel Window.}
\label{fig3} 
\end{center}
\end{figure}

After we have determined the Borel window, we 
can calculate the ground state mass, which is shown, as a 
function of $M^{2}_{B}$, in the Fig. \ref{fig3}. From this 
figure we see that there is a very good stability in the 
ground state mass in the determined Borel Window, which are
represented, through the crosses in Fig. \ref{fig3}. 

Varying the value of the continuum threshold in the 
range $\sqrt{s_{0}} = 4.70 \pm 0.10$ GeV, the mixing angle in the
range $\theta=(53.0\pm0.5)^0$, and the other
parameters as indicated in Table I, we get:
\begin{equation}
m_{Y} = (4.26 \pm 0.13) ~ \mbox{GeV},
\end{equation}
which is in a very good agreement with the experimental mass 
of the $Y(4260)$.

Once we have determined the mass, we can use this
value in Eq. (\ref{sumrule}) to estimate  
the meson-current coupling parameter, defined in
Eq. (\ref{coupling}). We have used the same values
of the $s_0$, $\theta$ and Borel Window used for the mass
calculation. Thus, we get:
\beq
\lambda_Y = (2.00 \pm 0.23) \times 10^{-2} ~ \mbox{GeV}^5.
\label{lay}
\enq
The  parameter $\lambda_Y$ gives a measure of the
strength of the coupling between the current and the state. 
The result in Eq.~(\ref{lay}) has the same order of magnitude 
as the coupling obtained for the $X(3872)$ \cite{x3872}, for example. 

%
\section{The Vertex function and the decay width of the $Y(4260)$}
%

The QCDSR technique can also be used to extract 
coupling constants and form factors. In particular, 
in Ref. \cite{bcnn} the authors determined the form 
factors and coupling constants in many hadronic vertices containing 
charmed mesons, in the framework of QCD sum rules. In this 
section, we will use the QCDSR approach to determine 
the coupling constant associated with the vertex $YJ/\psi \si$ 
to estimate the decay width of the process $Y \rightarrow J/\psi \pi \pi$. 
We are assuming that the two pions in the final state come
from the $\si$ meson. 
 
To determine the coupling constant  associated with the 
vertex $Y J/\psi \sigma$, we must evaluate the vertex 
function (three-point function) defined as
\beq
\Pi_{\mu \nu}(p,\pli, q) = \int d^4 x d^4y e^{i\pli \cdot x}
  e^{iq\cdot y}\Pi_{\mu \nu}(x,y),
\label{3po}
\enq
with $p=\pli+q$ and $\Pi_{\mu \nu}(x,y)$ given by
\beq
\Pi_{\mu \nu}(x,y)=\lag 0|T\{j_{\mu}^{\psi}(x)
j^{\sigma}(y)j_{\nu}^{Y\dagger}(0)\}|0\rag.
\label{pixy}
\enq
The interpolating fields appearing in Eq. (\ref{pixy}) 
are the currents for $J/\psi$, $\sigma$ and
$Y(4260)$, respectively. The currents for $J/\psi$ and 
$Y$ were defined by Eqs. (\ref{jcc}) and (\ref{jmix}). 
For the meson $\sigma$, we have
\beq
j^{\sigma}=\frac{1}{\sqrt{2}}\Big(\bar{u}_a(x)u_a(x)
+ \bar{d}_a(x)d_a(x)\Big).
\enq  

As in the case of two-point function studied in the 
previous section, the three-point correlation function 
defined by Eq. (\ref{3po}) can also be described in terms 
of hadronic degrees of freedom (Phenomenological side) or 
in terms of quarks and gluons fields (OPE side). In order 
to evaluate the phenomenological side of the sum rule we  
insert, in Eq.(\ref{3po}), intermediate states for $Y$, 
$J/\psi$ and $\sigma$. Using the definitions: 
\beq
\lag 0 | j_\mu^\psi|J/\psi(\pli)\rag =m_\psi 
f_{\psi}\epsilon_\mu(\pli), \nn
\enq
\beq 
\lag 0 | j^\sigma|\sigma(q)\rag =A_{\si}, \nn
\label{si}
\enq
\beq
\lag Y(p) | j_\nu^Y|0\rag =\la_Y \epsilon_\nu^*(p),
\lb{fp}
\enq
we obtain the following relation:
\beqa
\hspace{-1.0cm}\Pi_{\mu\nu}^{(phen)} (p,\pli,q)&=&{\la_Y m_{\psi}
f_{\psi} A_\sigma~ g_{Y\psi \sigma}(q^2)
\over (p^2-m_{Y}^2)({\pli}^2-m_{\psi}^2)(q^2-m_\sigma^2)} \\ \nn
&&\hspace{-1.0cm}\times((\pli \cdot p) g_{\mu\nu}
-\pli_\nu q_\mu - \pli_\nu \pli_\mu) +\cdots\;,
\lb{phen}
\enqa
where the dots stand for the contribution of all 
possible excited states. The form factor, $g_{Y\psi \si}(q^2)$, 
is defined by the generalization of the on-mass-shell
 matrix element, $\lag J/\psi \sigma|Y\rag$, for an off-shell 
$\sigma$ meson: 
\beq
\lag J/\psi\si |Y\rag=g_{Y\psi \sigma}(q^2)
(\pli \cdot p ~\epsilon^*(\pli)\cdot \epsilon(p) -
\pli \cdot \epsilon(p)~p\cdot \epsilon^*(\pli)),
\label{coup}
\enq
which can be extracted from the effective Lagrangian 
that describes the coupling between two vector mesons 
and one scalar meson:
\beq
{\cal{L}}=ig_{Y \psi \sigma}V_{\alpha\beta} A^{\alpha \beta}~\sigma
\enq
where $V_{\alpha \beta} = \partial_{\alpha} Y_{\beta} - 
\partial_{\beta} Y_{\alpha}$ and
$A^{\alpha \beta} = \partial^{\alpha} \psi^{\beta} - 
\partial^{\beta} \psi^{\alpha}$, are the tensor fields of the
$Y$ and $\psi$ fields respectively.

\begin{widetext}
\begin{center} 
\begin{figure}[h] 
\epsfig{figure=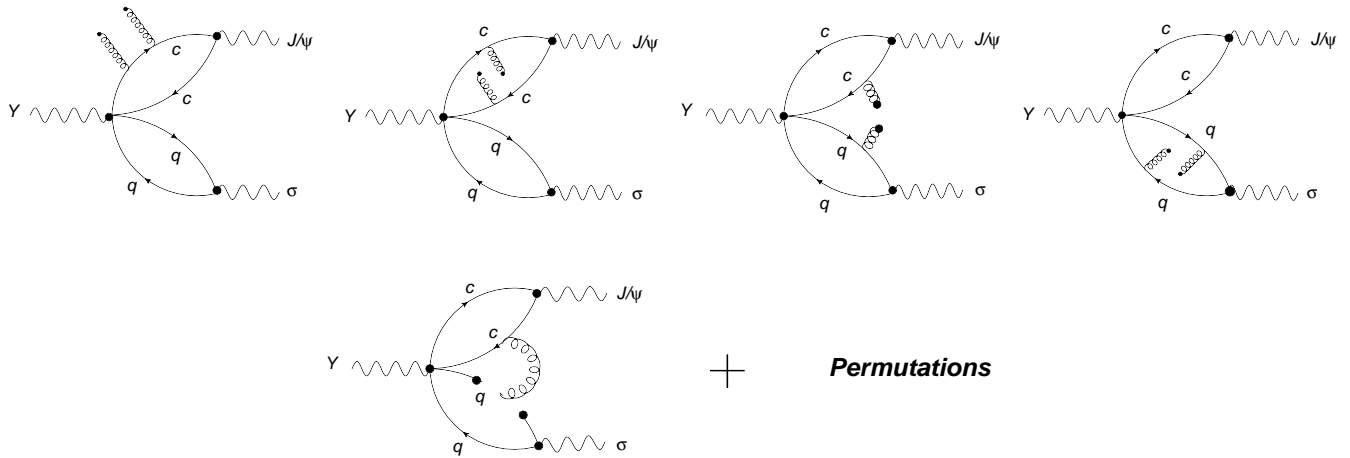,height=60mm}
\caption{\label{diags} Diagrams which contribute to the OPE side of the sum 
rule for the structure $\pli_{\nu}q_{\mu}$.}
\end{figure}   
\end{center}   
\end{widetext}

In the OPE side, we work at leading order in $\al_s$ and
we consider the condensates up to dimension five, as shown
in Fig. \ref{diags}. We have chosen to work in the 
${\pli}_{\nu} q_\mu$ structure since it has more 
terms contributing for the OPE. Taking the 
limit $p^2 = {\pli}^2=-P^2$ and doing the Borel transform 
to $P^2 \rightarrow M^2$, we get the following expression for 
the sum rule in the structure ${\pli}_{\nu} q_\mu$:
\beqa
\frac{\lambda_Y A_{\si} m_\psi f_\psi}
{(m_Y^2 -m_\psi^2)}~g_{Y\psi \sigma}(Q^2)\left(e^{-m_\psi^2/M^2}
-e^{-m_Y^2/M^2}\right)+\nn\\
+B(Q^2)~e^{-s_0/M^2}=(Q^2+m_\sigma^2)\Pi^{(OPE)}(M^2,Q^2),\nn\\
\label{3sr}
\enqa
where $Q^2=-q^2$, and $B(Q^2)$ gives the contribution to 
the pole-continuum transitions \cite{matheus,io2,decayx,dsdpi}.
$\Pi^{(OPE)}(M^2,Q^2)$ is given by
\beqa
\Pi^{(OPE)}(M^2,Q^2)&=&\frac{sin(\theta)}{3~2^4\sqrt{2}\pi^2}
\int\limits_{0}^{1} d\al e^{\frac{-m_c^2}{\al (1-\al)M^2}} \times \nn\\
&&\hspace{-3cm} \times \Big\{\frac{m_c\mix}{Q^2}
\Big[\frac{1-2\al(1-\al)}{\al(1-\al)}\Big]-\frac{\gG}{2^5\pi^4}\Big\}.\nn\\
\label{opeside}
\enqa

The sine present in Eq. (\ref{opeside}) indicates that
only the tetraquark part of current in Eq.~(\ref{jmix})
contributes to the OPE side. In fact, the charmonium part of the current
 gives only disconnected diagrams that are not considered. 

In Eq. (\ref{3sr})  $m_{\psi}$ and $f_{\psi}$ are the mass and decay constant of 
the $J/\psi$ and $m_{\si}$ is the mass of the $\si$ meson. Their values are: 
$m_{\psi}=3.1$ GeV, $f_{\psi}=0.405$ GeV \cite{pdg}, and 
$m_{\si}=0.478$ GeV \cite{ignacio}. The parameters $\la_{Y}$ and 
$A_{\si}$ represent, respectively, the coupling  of the $Y$ and $\si$ states 
with the currents defined 
in Eq. (\ref{coupling}) and (\ref{si}). The value of $\la_{Y}$ 
is given in Eq. (\ref{lay}), while $A_{\si}$ was 
determined in Ref. \cite{dosch} and its value is 
$A_{\si}=0.197$ GeV$^2$.

\begin{figure}[h]
\begin{center}
\centerline{\includegraphics[width=0.6\textwidth]{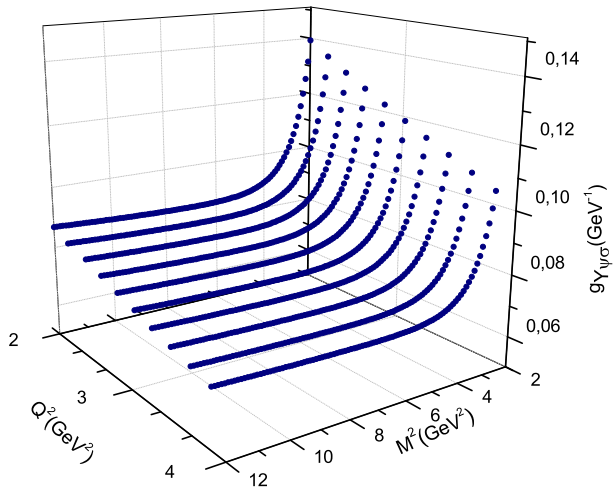}}
\caption{$g_{Y\psi\si}(Q^2)$ values obtained by varying both $Q^2$ and $M^2$.}
\label{fig3D} 
\end{center}
\end{figure}

Similarly to what was done to get $m_Y$ in Eq.~(\ref{mass}), one can
use Eq.~(\ref{3sr}) and its derivative with respect to $M^2$ to 
eliminate $B(Q^2)$ from these equations and to isolate $ g_{Y\psi \sigma}(Q^2)$. 
A good sum rule must be as much independent of the Borel mass
as possible. Therefore, we have to determine a region in
the Borel mass where the form factor is independent of $M^2$.
In Fig. \ref{fig3D} we show $g_{Y\psi\si}(Q^2)$ as a function of
both $M^2$ and $Q^2$. Notice that in the region 
$7.0 \leq M^2 \leq 10.0$ GeV$^2$, the form factor is clearly 
stable, as a function of $M^2$, for all values of $Q^2$.

\begin{figure}[h]
\begin{center}
\centerline{\includegraphics[width=0.6\textwidth]{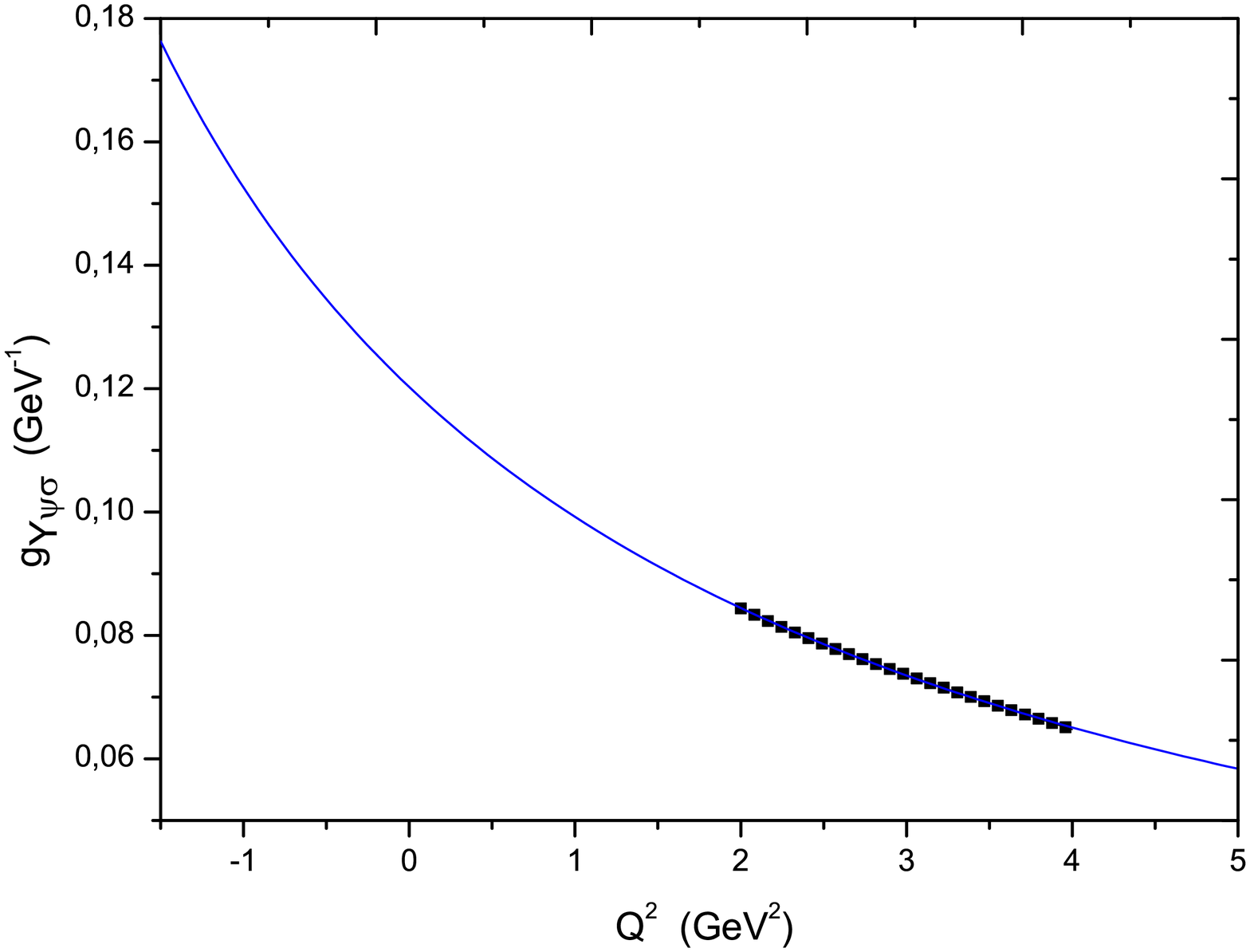}}
\caption{QCDSR results for $g_{Y\psi\si}(Q^2)$, as a function of $Q^2$, 
for $\sqrt{s_0}=4.76$ GeV (squares). 
The solid line gives the parametrization of the QCDSR results 
 through Eq. (\ref{mono}).}
\label{fig2D} 
\end{center}
\end{figure}

The squares in Fig. \ref{fig2D} show the $Q^2$ dependence 
of $g_{Y\psi\si}(Q^2)$, obtained for $M^2=8.0$ GeV$^2$. For 
other values of the Borel mass, in  the range  
$7.0 \leq M^2 \leq 10.0$ GeV$^2$, the results are equivalent.
Since we are interested in the coupling constant,  which is 
defined as value of the form factor at the meson pole: 
$Q^2 = -m^2_{\si}$,  we need to extrapolate 
the form factor for a region of $Q^2$ where
the QCDSR is not valid. This extrapolation 
can be done by parametrizing the QCDSR results for 
$g_{Y\psi\si}(Q^2)$ using a monopole form:

\beq
g_{Y\psi\si}(Q^2) = \frac{g_1}{g_2 + Q^2}.
\label{mono}
\enq

We do the fit for $\sqrt{s_0}=4.74$ GeV. We notice that the 
results do not depend much on this parameter. The results are:
\beq
g_1 = (0.58~\pm ~0.04)~\mbox{GeV}; ~~~ g_2=(4.71~\pm ~0.06)~\mbox{GeV}^2.  
\label{c1c2}
\enq

The solid line in Fig. \ref{fig2D} shows that the 
parametrization given by Eq. (\ref{mono}) reproduces very well the 
QCDSR results for $g_{Y\psi\si}(Q^2)$, in the interval $2.0 \leq Q^2 
\leq 4.0$ GeV$^2$, where the QCDSR is valid.

The coupling constant, $g_{Y\psi\si}$ is given by using 
$Q^2=-m^2_{\si}$ in Eq. (\ref{mono}). We get:

\beq
g_{Y\psi\sigma}=g_{Y\psi\sigma}(-m^2_\sigma)=(0.13 \pm 0.01)~~\mbox{GeV}^{-1}.
\label{coupvalue}
\enq
The error in the coupling constant  given above comes from 
variations in $s_0$ in the range $4.6\leq s_0 \leq 4.8$ 
GeV$^2$, and in the mixing angle $52.5^0\leq\theta\leq 53.5^0$.

In Table II, we show the other values of the coupling constant 
corresponding to the values of $\sqrt{s_0}$ that we have considered 
in our calculations.
\begin{center}
\small{{\bf Table II:} Monopole parametrization of the QCDSR results for
the chosen structure, for different values of $\sqrt{s_0}$}
\\
\vskip3mm
\begin{tabular}{ccc}  \hline
$\sqrt{s_0}~(\GeV)$&$g_{Y\psi\si}(Q^2)~(\GeV^{-1})$ & $g_{Y\psi\si}(Q^2=-m_\si^2)
~(\GeV^{-1})$\\
\hline
 4.6 & ${0.63\over Q^2+4.66}$ & 0.14 \\
 4.7 & ${0.57\over Q^2+4.71}$ & 0.13 \\
 4.8 & ${0.53\over Q^2+4.77}$ & 0.12 \\
\hline
\end{tabular}\end{center}

The decay width for the process 
$Y(4260) \rightarrow J/\psi \sigma \rightarrow J/\psi\pi \pi$
in the narrow width approximation is given by
\beqa
&&{d\Gamma\over ds}(Y\to J/\psi \pi\pi)={1\over8\pi m_Y^2}
|{\cal{M}}|^2\frac{m_Y^2 -m^2_{\psi}+s}{2m_Y^2} \nn\\
&\times&
{\Ga_{\si}(s) m_{\si}\over\pi}{p(s)\over(s-m_{\si}^2)^2+
(m_{\si}\Ga_{\si}(s))^2},
\label{de1}
\enqa
with $p(s)$ given by
\beq
p(s)={\sqrt{\la(m_Y^2,m_\psi^2,s)}\over2m_Y},
\enq
where $\la(a,b,c)=a^2+b^2+c^2-2ab-2ac-2bc$, and $\Ga_{\si}(s)$ 
is the s-dependent width of an off-shell $\sigma$ meson \cite{ignacio}:

\beq
\Ga_{\si}(s)=\Ga_{0\si}\sqrt{\frac{\la(s,m_{\pi}^2, m_{\pi}^2)}
{\la(m_Y^2,m_{\pi}^2, m_{\pi}^2)}}\frac{m_Y^2}{s},
\enq
where $\Ga_{0\si}$ is the experimental value for the decay 
of the $\si$ meson into two pions. Its value is 
$\Ga_{0\si}=(0.324\pm 0.042\pm 0.021)$ GeV \cite{ignacio}.

The invariant amplitude squared can be obtained from 
the matrix element in Eq. (\ref{coup}). We get:
\beq
|{\cal M}|^2=g_{Y\psi \sigma}^2(s)f(m_Y,m_\psi,s),
\enq
where $g_{Y\psi \si}(s)$ is the form factor in the 
vertex $YJ/\psi \si$, given in Eq.~(\ref{mono}) using $s=-Q^2$, and
\beqa
&&f(m_Y,m_\psi,s)={1\over3}\left(m_Y^2m_\psi^2 + 
{1\over2}(m_Y^2 + m_\psi^2 - s)^2\right). \nn
\label{m2}
\enqa

Therefore, the decay width for the process 
$Y(4260)\rightarrow J/\psi \pi\pi$ is 
given by
\beq
\Ga = \frac{m_\si}{16\pi^2 m^4_Y}I,
\label{width}
\enq
where we have defined
\beqa
I &=& \int_{(2m_\pi)^2}^{(m_Y-m_\psi)^2} ds~g_{Y\psi\si}^2(s)
\Ga_\si(s)(m^2_Y-m^2_\psi +s)   \nn \\
&\times & f(m_Y,m_\psi, s) \frac{p(s)}{(s-m^2_\si)^2+(m_\si \Ga_\si(s))^2}.
\enqa
Hence, taking variations on $s_0$ and $\theta$ in the same intervals 
given above, we obtain from Eqs. (\ref{coupvalue})-(\ref{width}) the 
following value for the decay width 

\beq
\Ga(Y\rightarrow J/\psi \pi\pi) = (1.0\pm 0.2 ) ~\mbox{MeV},
\enq
which is not compatible with the experimental decay width value 
expected for the $Y(4260)$ state which is around 
$\Ga_{exp}\approx (88~\pm~23)$ MeV \cite{babar1}.

%
\section{Summary and Conclusions}
%

In summary, we have used the QCDSR approach to study the two-point and 
three-point functions of the $Y(4260)$ state, by considering a 
mixed charmonium-tetraquark current.  In the determination of the
mass, we work with the two-point function at leading order in 
$\alpha_{s}$ and we consider the contributions from the
condensates up to dimension eight. A very good agreement with the
experimental value of the mass of the $Y(4260)$
is obtained for the mixing angle around $\theta \approx (53.0\pm 0.5)^{0}$. 

To evaluate the width of the decay $Y(4260)\to J/\psi\pi\pi$, we work 
with the three-point function also at leading order in 
$\alpha_{s}$ and we consider the contributions from the
condensates up to dimension five. We assume that the two pions in the final
 state come
from a $\sigma$ meson. The obtained value for width is
$\Ga_Y \approx (1.0\pm 0.2)$ MeV, which is much smaller than the
experimental value: $\Ga_{exp} \approx (88\pm 23)$ MeV.

Therefore, we conclude that the $Y(4260)$ exotic state cannot be
described as a mixed charmonium-tetraquark state.    

%
\subsection*{Acknowledgment} 
%

This work has been supported by FAPESP and CNPq.

\appendix
\label{apa}
%
\section{The spectral densities for charmonium and tetraquark}
%

Next, we list all the spectral densities that appear in
Eqs. (\ref{pi22})-(\ref{pi24}) for charmonium 
$\Pi_1^{22}(M_B^2)$, tetraquark 
$\Pi_1^{44}(M_B^2)$ state as well as the mixed terms 
$\Pi_1^{24}(M_B^2)$ and $\Pi_1^{42}(M_B^2)$. 
The contributions for the last two are equal, that is, 
$\Pi_1^{24}(M_B^2)=\Pi_1^{42}(M_B^2)$.

For the charmonium contribution, the spectral densities are
written below \cite{rry}
\beqa
  \rho^{pert}_{22}(s) &=& 
  \frac{s \qq^2}{2^3 \pi ^2} (1+2m_c^2/s) \sqrt{1 - 4m_c^2/s},
  \nn\\&&\\ 
  \Pi^{\GG}_{22}(M_B^2) &=& -\frac{\gG \qq^2}{3 \cdot 2^6 \pi^2} 
   \int\limits^{1}_{0} \!d\al \Big\{2+ \nn\\
&&+\hspace{-2.8cm} \frac{m_c^2(1-7\alpha -2 \alpha ^2)}
{\alpha (1-\alpha )^2M_B^2}+
  \frac{4m_c^4}{M_B^4(1-\alpha )^3}\Big\}e^{-\frac{m_c^2}
  {M_B^2\alpha (1-\alpha )}}.\nn\\
\enqa

For the tetraquark we have
\beqa
\rho^{pert}_{44}(s) &=& -\frac{1}
{3\cdot 2^{10}\pi ^6}
\int\limits^{\almax}_{\almin} \!\!\frac{d\al}{\alpha^3} 
\int\limits^{\bemax}_{\bemin} \!\!\frac{d\be}{\beta^3} 
F^3(1-\alpha -\beta ) \nn \\
&&\hspace{-1cm} \times \Bigg( 2m_c^2 (1-\alpha -\beta )^2 - 
3F (1+\alpha +\beta ) \Bigg),\\
\rho^{\qq}_{44}(s) = 0 
\enqa
\beqa
\rho^{\GG}_{44}(s) &=& -\frac{\gG}
{3^2 \cdot 2^{11} \pi^6}
\int\limits^{\almax}_{\almin} \!\!\frac{d\al}{\al}
 \int\limits^{\bemax}_{\bemin}
\!\!\frac{d\be}{\beta ^3}\times \nn\\
&&\hspace{-2.0cm}\times\bigg[ 2 m_c^4 \alpha 
(1-\alpha -\beta )^3 
- 3m_c^2 F(1-\alpha -\beta)\times\nn\\
&&\hspace{-2.0cm}\times \left(2\alpha ^2+\alpha 
(8+3\beta )+\beta (1+\beta )-2\right)+\nn\\
&&+6 F^2 \beta  (1-2 \alpha -2 \beta )\bigg],\\
\rho^{\qGq}_{44}(s) &=& -\frac{\qGq}
{3 \cdot 2^7 \pi^4} \Bigg\{ 
3m_c \int\limits^{\almax}_{\almin} \!\!\frac{d\al}{\al^2}
\int\limits^{\bemax}_{\bemin} \!\!\frac{d\be}{\be}
F \bigg[ \alpha^2 -\nn\\
&&\hspace{-2.0cm} - \alpha  (1+\beta) - 2 \beta^2 \bigg]
+ m_s \int\limits^{\almax}_{\almin} \!\!d\al \Bigg[ 16m_c^2 +
2H \left( \frac{1-\al}{\al} \right) -\nn\\
&&\hspace{-2.0cm}- \int\limits^{\bemax}_{\bemin} 
\!\!\frac{d\be}{\be} \bigg( m_c^2(9-3\al-5\be)
+ 7F \bigg) \Bigg] \Bigg\},
\enqa

\beqa
\rho^{\qq^2}_{44}(s) &=& \frac{s \qq^2}
{3^2 \cdot 2^4 \pi ^2} (1-16m_c^2/s) \sqrt{1- 4m_c^2/s} \\
\rho^{\langle 8 \rangle}_{44}(s) &=& -\frac{\qq \qGq}
{3 \cdot 2^5 \pi^2} \int\limits^{\almax}_{\almin} 
\!\!d\al \:\alpha(5-6 \alpha )\\
\Pi^{\lag 8 \rag}_{44}(M_B^2) &=& 
-\frac{m_c^2 \qq \qGq}{3 \cdot 2^4 \pi^2}
\int\limits^{1}_{0} \!d\al \times \nn\\
&&\times \Bigg[ \frac{\alpha^2
- 2m_c^2}{M_B^2\al(1-\al)} \Bigg]
 ~e^{-\frac{m_c^2}{M_B^2\al (1-\al)}},  
\enqa
Finally, for the mixed term we have
\beqa
\rho^{\qq}_{24}(s) &=& -\frac{s \qq^2}
{3 \cdot 2^3 \pi ^2}
 (1+2m_c^2/s) \sqrt{1 - 4m_c^2/s},\nn\\&&\\
\Pi^{\qGq}_{24}(M_B^2) &=&
 -\frac{m_c^2 \qq \qGq}
{3 \cdot 2^3 \pi^2} \int\limits^{1}_{0} \! \frac{d\al}{\al}
~e^{-\frac{m_c^2}{M_B^2\al (1-\al)}}\nn \\
\enqa

In all these expressions we have used the following definitions:
\beqa
F=(\al + \beta)m_c^2-\al \beta s,\\
H=m_c^2-\al(1-\al)s,
\enqa

and the integration limits are:
\beqa
\al_{min} =\frac{1-\sqrt{1-4m_c^2/s}}{2},\\
\al_{max}=\frac{1+\sqrt{1-4m_c^2/s}}{2},\\
\beta_{min}=\frac{\al m_c^2}{(s\al-m_c^2)}.
\enqa

\end{document}